\documentclass{naturemod}
\usepackage{graphicx}
\usepackage{graphics}
\usepackage{amssymb}
\usepackage{hyperref}
\usepackage{xcolor}
\usepackage{amsmath}
\usepackage{newtxtext}
\usepackage[varg,smallerops,upint]{newtxmath}
\usepackage{newfloat}

\DeclareFloatingEnvironment[fileext=lop]{Extended Data Figure}

\title{A dynamically cold disk galaxy in the early Universe}

\author{F. Rizzo$^{1}$, S. Vegetti$^{1}$, D. Powell$^{1}$, F. Fraternali$^{2}$, J. P. McKean$^{2,3}$, H. R. Stacey$^{2,3,1}$ \& S. D. M. White$^{1}$}

\begin{document}

\maketitle

\begin{spacing}{1.2}
\begin{affiliations}
\item Max-Planck Institute for Astrophysics, Karl-Schwarzschild Str 1, D-85748 Garching, Germany 
\item Kapteyn Astronomical Institute, University of Groningen, Postbus 800, NL-9700 AV Groningen, the Netherlands
\item ASTRON, Netherlands Institute for Radio Astronomy, Oude Hoogeveensedijk 4, NL-7991 PD Dwingeloo, the Netherlands
\end{affiliations}

\begin{abstract}
The extreme astrophysical processes and conditions that characterize the early Universe are expected to result in young galaxies that are dynamically different from those observed today\cite{pillepich, dekel, zolotov15, krum, hay}. This is because the strong effects associated with galaxy mergers and supernova explosions would lead to most young star-forming galaxies being dynamically hot, chaotic and strongly unstable\cite{pillepich, dekel}. Here we report the presence of a dynamically cold, but highly star-forming, rotating disk in a galaxy at redshift ($z$) 4.2\cite{weis}, when the Universe was just 1.4 billion years old. Galaxy SPT--S J041839--4751.9 is strongly gravitationally lensed by a foreground galaxy at $z = 0.263$, and it is a typical dusty starburst, with global star-forming\cite{deb} and dust properties\cite{gullberg} that are in agreement with current numerical simulations\cite{mcalpine} and observations of its galaxy population\cite{hodge}. Interferometric imaging at a spatial resolution of about 60 pc reveals a ratio of rotational-to-random motions of $V/\sigma = 9.7\pm 0.4$, which is at least four times larger than expected from any galaxy evolution model at this epoch\cite{pillepich, dekel, zolotov15, krum, hay}, but similar to the ratios of spiral galaxies in the local Universe\cite{lelli16}. We derive a rotation curve with the typical shape of nearby massive spiral galaxies, which demonstrates that at least some young galaxies are dynamically akin to those observed in the local Universe, and only weakly affected by extreme physical processes.
\end{abstract}

\begin{figure}[h!]
	\centering
	\includegraphics[width=1.03\textwidth]{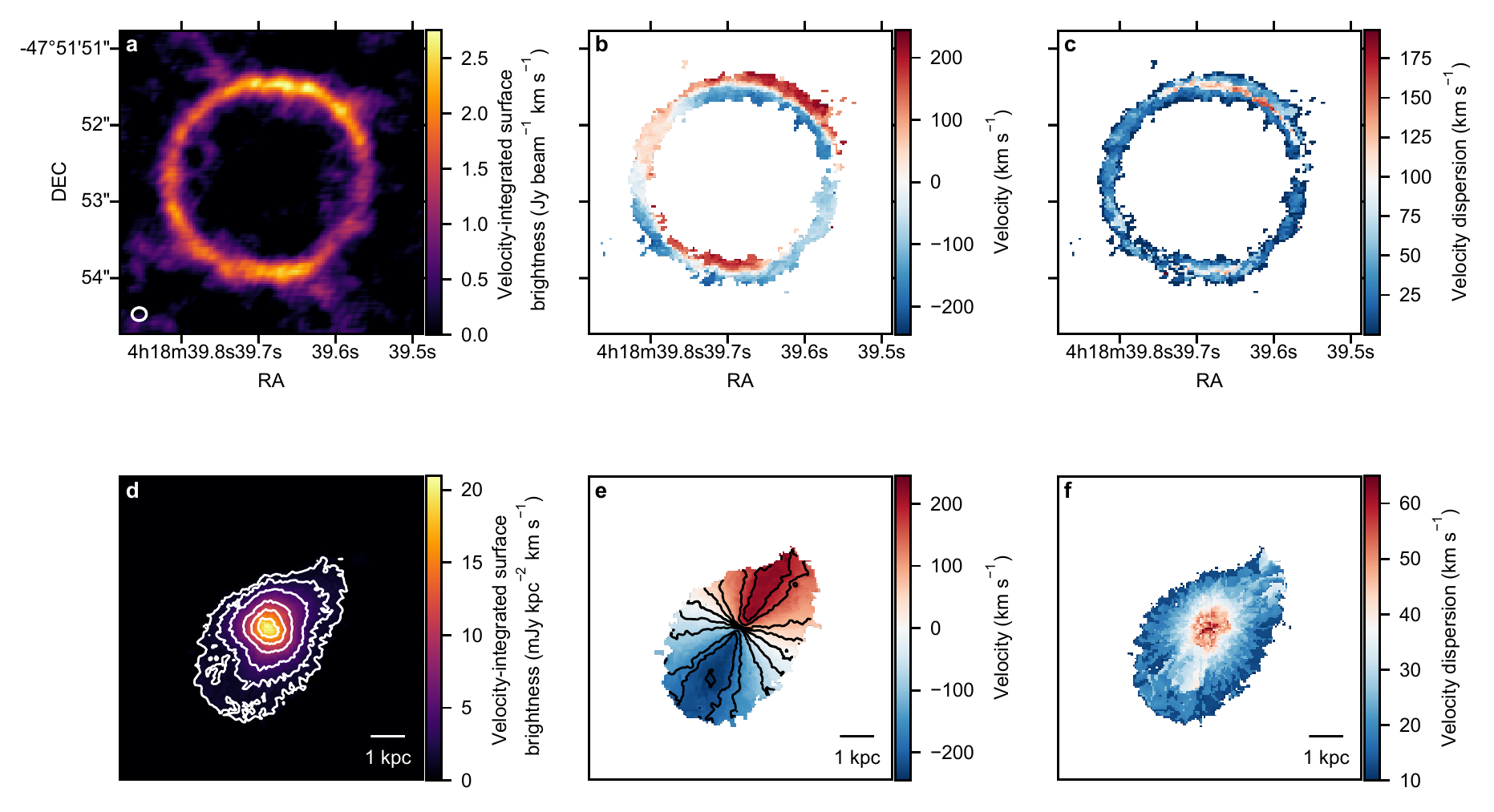}
	\caption{\small{\textbf{[CII] emission from the lensed galaxy SPT0418-47 and source plane reconstruction.} Panel a: emission of the 158-$\mu$m fine-structure 	line of ionized carbon [C II] integrated across a velocity range of 721 km s$^{-1}$ (zeroth-moment map). The beam size, shown as a white ellipse on the lower 	left corner, is $0.19" \times 0.17"$ at a position angle of -85.22$^{\circ}$. Panels b and c: same as in panel a, but the emission is colour-coded by the flux-weighted velocity and velocity dispersion (first- and second-moment maps) respectively. Panels d-f: zeroth-, first- and second-moment maps of the reconstructed source. In panel d the white contours are set at n = 0.05, 0.1, 0.2, 0.4, 0.6, 0.8 times the value of the maximum flux of the zeroth-moment map. In Panel e the black solid contours are at $V_{\mathrm{sys}}\pm \Delta V$, where $V_{\mathrm{sys}}$ is the systemic velocity set to 0 km s$^{-1}$ and $\Delta V = 40$ km s$^{-1}$. We note that the scales of the second-moment maps (panels c and f) are different because panel c shows the observed values, while panel f shows the intrinsic ones (beam-smearing corrected). These six maps are intended only for visualisation, as the source reconstruction and its kinematic modelling are performed using the full 3D information of the data cube containing the [CII] emission line (see Methods and Extended Data Fig. \ref{ed1} for further details).}}
	\label{fig1}
\end{figure}

\begin{figure}
	\centering
	\includegraphics[width=1.01\textwidth]{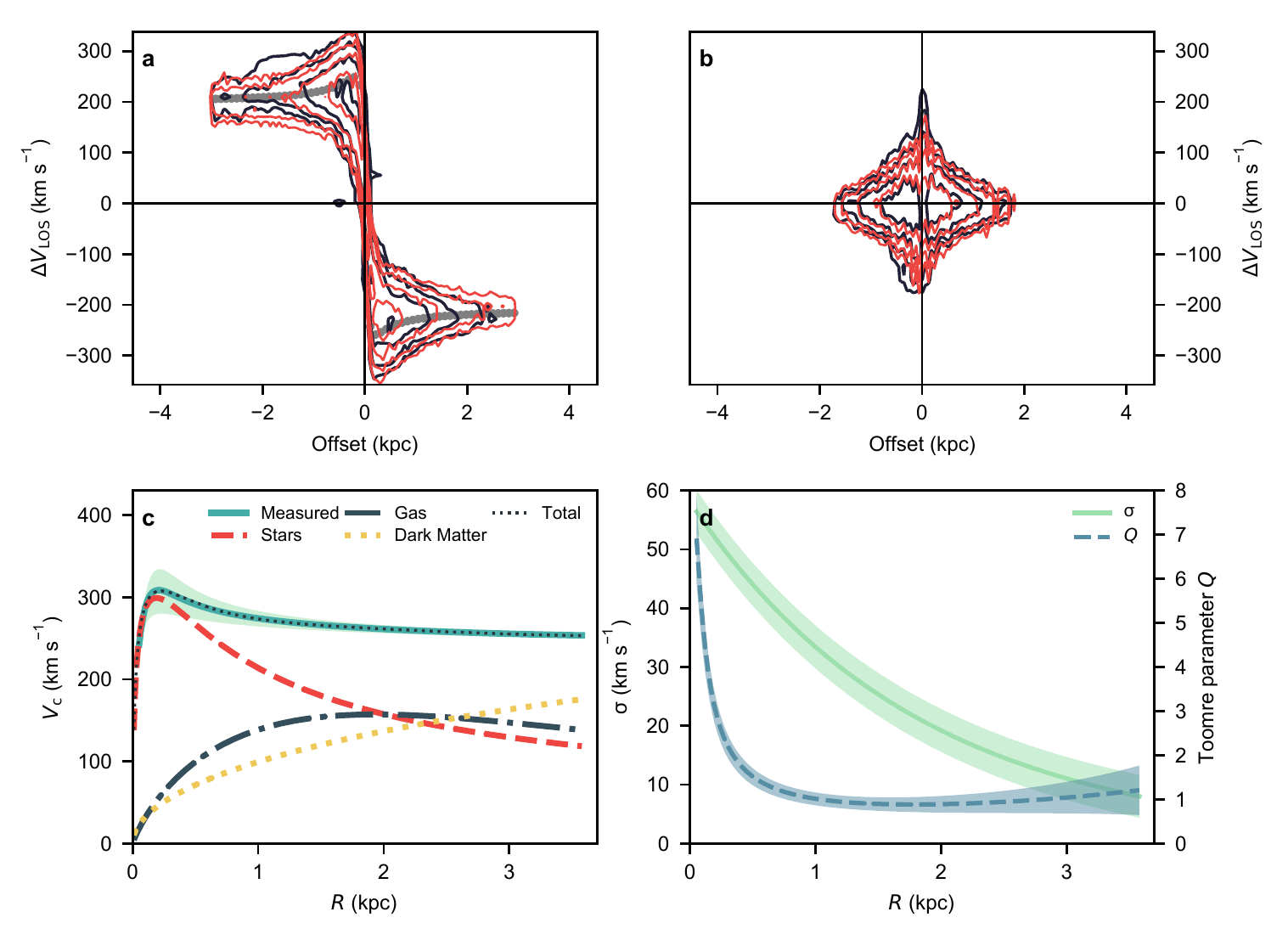} 
	\caption{\small{\textbf{Kinematic and dynamical properties of SPT0418-47.} Panel a and b: [CII] emission in the position-velocity diagrams along the major and minor axis respectively. These diagrams show slices, the equivalent of putting long slits along the two axes.  The x axis shows the offset along the major and minor axis from the galaxy center, and the y axis represents the line-of-sight velocity centred at the systemic velocity of the galaxy. The dark blue contours show the reconstructed source, and the red contours show the best kinematic model (see Methods). The contour levels are set at $n = 0.1, 0.2, 0.4, 0.8$ times the value of the maximum flux in the major-axis position-velocity diagram. The grey circles show the rotation velocities derived using our 3D lens-kinematic methodology. Panel c: rotation curve decomposition. The green solid line shows the circular velocity profile. The black dotted line is the best dynamical model, obtained by fitting the different mass components contributing to the total gravitational potential as shown in the legend. Panel d: velocity dispersion profile (solid green line) and Toomre parameter profile (dotted blue line). The coloured bands in panel c and d represent uncertainties obtained by error propagation from the 1 s.d. uncertainties of the parameters that define the respective profiles (see Methods and Extended Data Table \ref{tab1}).}}
	\label{fig2}
\end{figure}

SPT--S J041839--4751.9 (hereafter SPT0418--47) has been observed with the Atacama Large Millimeter/submillimeter Array (ALMA) to image the thermal continuum emission from the dust and the emission from the 158-$\mu$m fine-structure line of ionized carbon [CII] (Fig. \ref{fig1}a). The [CII] line is typically the brightest fine-structure line emitted in star-forming galaxies, and is a main gas coolant, coming from multiple phases of the inter-stellar medium (ISM), including the warm ionized, the warm and cold neutral atomic, and the dense molecular medium\cite{stacey}. We applied a three-dimensional (3D) lens-kinematic modelling technique\cite{rizzo2} to the interferometric data cube containing the [CII] line to determine the gas surface brightness distribution in each spectral channel (Fig. \ref{fig1}, \ref{fig2} and Extended Data Figs.\,\ref{ed1}, \ref{ed2} and Table \ref{tab1}; see Methods for full discussion), from which we derive the gas kinematics, and perform a robust dynamical analysis of the galaxy. In addition, we reconstructed the far-infrared surface brightness distribution of the heated dust emission from the interferometric continuum dataset.

We find that the rotation curve of SPT0418--47 has the typical shape of a bulge-domin.ated spiral galaxy in the local Universe\cite{lelli16}; it has a bump at 0.2 kpc from the galaxy centre and then declines before flattening at radii larger than 1.5~kpc (Figs.\,\ref{fig2}a, c). The [CII] velocity dispersion $\sigma$ is well described by an exponential profile, with average values of $\sim$ 45 km\,s$^{-1}$ in the inner regions ($\lesssim 1$~kpc) and $\sim$18 km\,s$^{-1}$ in the outer disk ($\gtrsim1$~kpc) (Figs. \ref{fig2}b, d, Extended Data Table \ref{tab1}). From a decomposition of the rotation curve, we derive the relative contribution of the different mass components to the total galactic gravitational potential, the stellar component, the gaseous disk and the dark matter halo (Extended Data Table \ref{tab5}, Fig. \ref{fig2}c; see Methods). \\
SPT0418--47 has global physical properties (total stellar mass $M_{\mathrm{star}} = 1.2^{+0.2}_{-0.1} \times 10^{10}$~M$_{\odot}$, dark matter mass $M_{\mathrm{DM}} = 1.7^{+0.3}_{-0.3} \times 10^{12}$~M$_{\odot}$, gas fraction $f_{\mathrm{gas}} = 0.53^{+0.06}_{-0.08}$, and stellar effective radius $R_{\mathrm{e}} = 0.22^{+0.04}_{-0.02}$ kpc; see Extended Data Table \ref{tab5} for further parameters) that are in agreement with the predictions from the most recent theoretical models\cite{mcalpine}, as well as observations of the population of starburst galaxies at this redshift\cite{hodge}. From the de-lensed dust emission, we derive an intrinsic infrared luminosity of ($2.4\pm0.4) \times10^{12}$~L$_{\odot}$, a star formation rate (SFR) of $352\pm65$ M$_{\odot}$~yr$^{-1}$ and a gas depletion timescale of $38\pm9$~Myr, which are all typical of mm-selected starburst galaxies at this redshift\cite{hodge} (see Methods).

\begin{figure}
	\centering
	\includegraphics[width=0.85\textwidth]{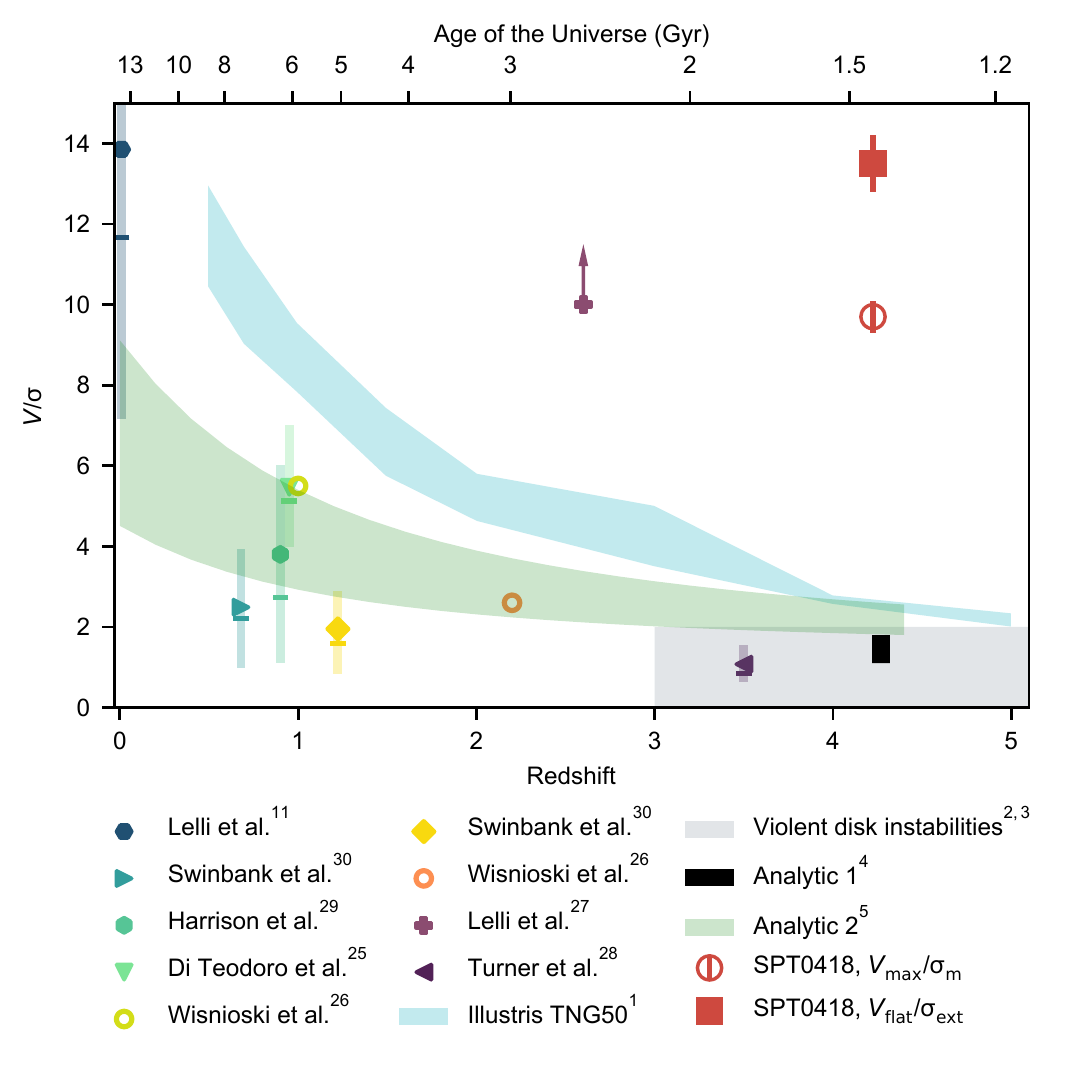}
	\caption{\small{\textbf{Comparison between SPT0418-47 and samples of observed and simulated galaxies.} Mean ratios of the rotational to random motions ($V/\sigma$) versus redshifts for the comparison samples\cite{lelli16, edt, wisni, lelli, turner, harrison, swin17} of observed star-forming galaxies indicated in the legend and for SPT0418-47 (red square and red empty circle). The gas tracers are: HI\cite{lelli16}, H$\alpha$\cite{edt, wisni, harrison, swin17}, [OII]\cite{swin17}, [OIII]\cite{turner}, [CI]\cite{lelli}. For SPT0418-47,  the $V/\sigma$ was obtained from the [CII] emission line. For the comparison samples, the shaded regions show the area between the 16th and 84th percentiles of the distributions, while the horizontal bars show the median values (when available). In Extended Data Table \ref{tab3} we show the different extraction methods used to compute $V/\sigma$. For the empty markers the $V/\sigma$ values were calculated using for each galaxy a proxy for the maximum rotation velocity, $V_{\mathrm{max}}$. The $V/\sigma$ values shown by the full markers were calculated using the flat or the outer part of the rotation curve $V_{\mathrm{flat}}$. The violet cross is a lower limit for a single galaxy\cite{lelli}. The light-blue band shows the $V/\sigma$ for simulated galaxies from TNG50 simulation \cite{pillepich} in the mass range $10^9 - 10^{11} M_{\odot}$. For these simulated galaxies, $V/\sigma$ is calculated as the ratio between the maximum rotation velocity, $V_{\mathrm{max}}$ and the mean velocity dispersion $\sigma_{\mathrm{m}}$. The grey area show the expected $V/\sigma$ for galaxies dominated by violent disk instabilities\cite{zolotov15, dekel}. The black and green areas show a prediction and an assumption, respectively, from two different analytical models (Analytic 1\cite{krum} and Analytic 2\cite{hay}). The red square is the $V_{\mathrm{max}}/\sigma_{\mathrm{m}}$ for SPT0418-47, and the empty red circle shows the position of the $V/\sigma$ value for SPT0418-47, calculated as the ratio between $V_{\mathrm{flat}}$ and the velocity dispersion at outer radii $\sigma_{\mathrm{ext}}$ (see Extended Data Table \ref{tab2}). The error bars for the red square and empty circles show the 1 s.d. uncertainties.
	}}
	\label{fig3}
\end{figure}

Our high-resolution 3D kinematical analysis shows that SPT0418--47 has a ratio of rotational velocity ($V$) to velocity dispersion $\sigma$ of $V/\sigma = 9.7\pm0.4$ (Extended Data Table \ref{tab2}), which is similar to that of spiral galaxies in the local Universe\cite{lelli16}, but considerably larger than what is predicted by most numerical\cite{pillepich, zolotov15} and analytical\cite{dekel, krum, hay} models. For example, for the majority of star-forming galaxies at $z \approx 4$ with stellar masses in the range $10^{9}$--$10^{10.5}$~M$_{\odot}$, the most recent cosmological magnetohydrodynamical simulation TNG50\cite{pillepich} gives $V/\sigma \lesssim 3$ (light-blue band in Fig. \ref{fig3}); even though such simulated galaxies have rotationally supported gas-rich disks, they are dynamically hotter than their low-redshift counterparts. Complex astrophysical processes, such as stellar feedback by supernovae or radiation pressure, or active galactic nucleus (AGN) feedback from the central supermassive black-hole, galaxy mergers, and gas inflows and outflows, are expected to have a significant impact on the gas kinematics within galaxies at this early epoch, and are predicted to be responsible for a progressive increase of chaotic, random motions with redshift\cite{pillepich, krum, hay}. Our finding firmly rules out models in which a high star-formation feedback and a high gas fraction necessarily produce large turbulent motions\cite{pillepich, krum} (light-blue, green, and black bands in Fig. \ref{fig3}) and violent disk instabilities\cite{zolotov15, dekel}, resulting in dispersion dominated systems with $V/\sigma \lesssim 2$ at these redshifts (grey band in Fig. \ref{fig3}; see Methods for further discussions about these comparisons). Our result requires galaxy evolution models to produce dynamically cold galaxies that are not characterised by large turbulent motions\cite{pillepich, krum} and violent instabilities\cite{dekel, zolotov15}, already at early times.

Our kinematic analysis also allows the level of axisymmetric disk instabilities to be measured within SPT0418--47 via the Toomre parameter\cite{toomre} $Q$: a value of $Q > 1$ ensures that no instabilities will develop because large-scale collapse is prevented by differential rotation, whereas $Q \lesssim 1$ indicates that instabilities will be able to grow and lead to the formation of gas and star-formation clumps within the disk. For SPT0418--47, we find an average value of $Q = 0.97\pm 0.06$ at $R > 1$~kpc (Fig. \ref{fig2}d), where the gas component is dominant (Fig. \ref{fig2}c), indicating a potentially unstable disk, prone to form clumpy star-forming regions. This result supports the hypothesis\cite{carilli} that the irregular morphologies found for dusty starburst galaxies in the optical/ultraviolet rest-frame wavelengths\cite{chen} are poorly resolved clumpy star-forming regions, and not objects that are undergoing or have recently experienced a merging event. 

Using our rotation curve decomposition (see Methods), we find that the stellar component of SPT0418--47 is well described by a Sérsic profile with a Sérsic index of $2.2^{+0.3}_{-0.2}$ and a stellar mass of $1.2^{+0.2}_{-0.1} \times10^{10}$~M$_{\odot}$. Several observational studies of the structural properties\cite{lang} of galaxies have confirmed that the Hubble sequence is already in place at $z \approx 2.5$, with galaxies showing a large variety of morphologies. However, at $z$ $\gtrsim 3$, the lack of spatially resolved data in the rest-frame optical emission for these galaxies has prevented the study of their structure and morphologies. The unprecedented spatial resolution of 60 pc of the dataset for SPT0418--47 allows us to study, for the first time, the morphological properties of a $z$ $\approx 4$ galaxy. The bump in the inner region of the rotation curve clearly indicates that a bulge is already in place at $z$ $\approx 4$, whereas the Sérsic index of $\sim 2$ is a signature of either a disky bulge\cite{kraj} or an embedded disk-like component.

\begin{figure}
	\centering
	\includegraphics[width=1.0\textwidth]{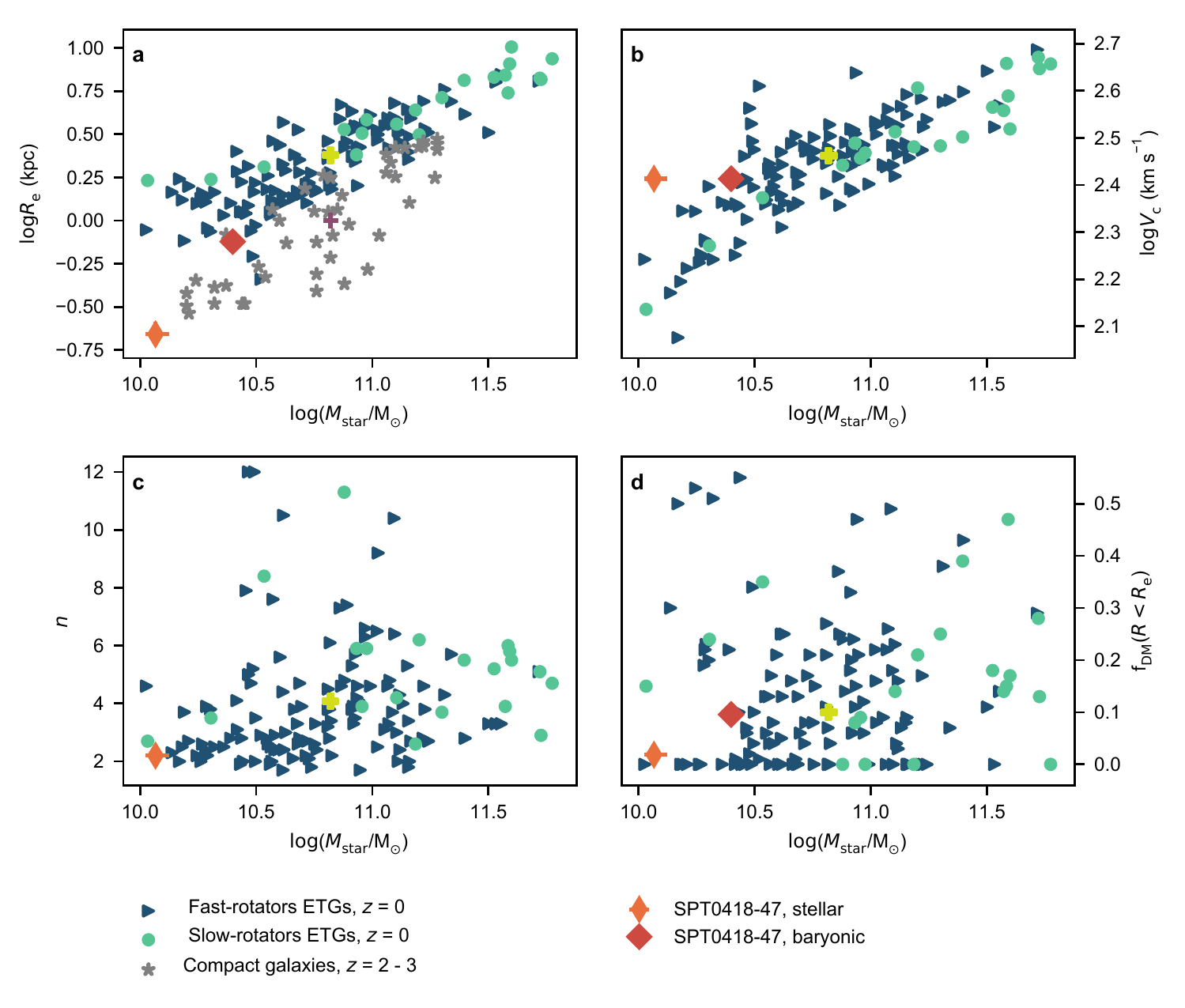}
	\caption{\small{\textbf{Comparison between SPT0418-47 and samples of its plausible descendants.} In all panels, the orange thin diamond shows the position of SPT0418-47, and the red diamond shows the baryonic quantities (gas plus stars), under the assumption that all the gas that we observe today will be converted into stars, preserving the disk configuration. The error bars show the 1 s.d. uncertainties. The z $\approx$ 0 ETGs (from the ATLAS$^{\mathrm{3D}}$ survey\cite{cappellari, mcdermid, kraj, cappellari2013b}), the plausible descendants of SPT0418-47, are shown according to their kinematic classification: blue triangles show fast rotators, green circles show slow rotators and yellow crosses show the median values. Panel a: size-stellar mass plane. The grey stars show the compact galaxies\cite{barro} at z $\approx$ 2-3 and the violet cross show the median values for this sample. Panel b: circular velocity versus stellar mass. Panel c: Sérsic index versus stellar mass. Panel d: fraction of dark matter within the effective radius versus stellar mass.}}
	\label{fig4}
\end{figure}

Dusty starburst galaxies are believed to be the progenitors of early-type galaxies (ETGs), which are the most massive galaxies observed today, dominated by old stellar populations. The most popular evolutionary track for this transformation\cite{toft, zolotov15, dekel} predicts that the dusty-starburst phase is followed by a quenching phase, during which AGN feedback leads to gas consumption and heating with the consequent formation of a population of compact quiescent galaxies\cite{barro} at $z$ $\approx 2$. In the final phase, dry minor mergers are expected to be responsible for a growth in galaxy size and the transformation into present-day ETGs. In Fig. \ref{fig4}a-d we compare the main physical quantities of SPT0418--47, as inferred from our dynamical model, with the corresponding quantities for the sample of local ETGs from the ATLAS$^{\mathrm{3D}}$ survey\cite{cappellari}. We consider only those local ETGs with estimated stellar ages $\gtrsim 12$~Gyr\cite{mcdermid}, which is the lookback time corresponding to $z$ $\approx 4$. The orange thin diamond in each panel shows the values derived for SPT0418--47 as observed today, and the red diamond shows the corresponding baryonic values (gas plus stars), under the assumption that all the gas that we observe today will be converted into stars, preserving the disk configuration. Given the SFR estimated for SPT0418--47, this conversion will happen in $\sim 38$~Myr (see Methods). \\
The comparison between the ETGs and the stellar/baryonic quantities for SPT0418--47 in the size--mass plane (Fig. \ref{fig4}a) indicates that this starburst galaxy should increase its stellar mass by a factor of 6 (3 for the red diamond) and its effective radius by a factor of 11 (3 for the red diamond), in order to evolve into an average ETG (yellow cross). This is in agreement with a simple toy model\cite{naab} for mergers, in which a single dry major merger event would be responsible for an increase in both the size and stellar mass of SPT0418--47 by a factor of 3. In Fig. \ref{fig4}a, the grey stars show the populations of compact galaxies\cite{barro} at $2 < z < 3$. Notably, the median size for this population is comparable to the size of SPT0418--47 (red diamond), but its median stellar mass is a factor of 3 larger. Also, the position of SPT0418--47 on the mass--size plane is  compatible with the low-mass end of both ETGs and $z$ $\approx 2$ compact galaxies. This implies that either this galaxy will evolve smoothly into a low-mass ETG after the consumption and/or heating of the cold gas reservoir or, as predicted by the merger scenario, it will reach the bulk of the ETG population in the size--mass plane. Fig. \ref{fig4}c shows that a potential progenitor of an ETG has already at $z \approx$ 4 a disky stellar component, a feature that is very common especially amongst local fast rotator ETGs\cite{kraj}. Finally, we derived the fraction of dark-matter mass within the effective radius and found that, with a value of $f_{\mathrm{DM}}(<R_{\mathrm{e}})=0.095^{+0.004}_{-0.004}$ (red diamond), the central regions of SPT0418--47 are dominated by baryons. As shown in Fig. \ref{fig4}c, such a low fraction of dark matter is compatible with observations of local ETGs\cite{cappellari2013b}, implying that the physical mechanisms responsible for the mass and size growth of this galaxy with cosmic time should preserve the dark-matter contribution within the innermost $\sim 1$ kpc.

\clearpage
\begin{methods}
\subsection{Observations.}
SPT0418--47 was identified during the South Pole Telescope Survey\cite{carlstrom} as a far-infrared luminous source with dust-like spectral indices at 1.4 and 2.0~mm\cite{vieira}. Follow-up observations\cite{vieira} with ALMA confirmed SPT0418--47 to be a dusty starburst galaxy at redshift ($z$) 4.2248\cite{weis} that is strongly gravitationally lensed by a foreground galaxy at $z = 0.263$. \\
The data used for the analysis described here were taken from the ALMA Archive. SPT0418--47 was observed with ALMA on 5 July 2017 under project code 2016.1.01499.S (PI:~K. Litke). Observations were taken at a central frequency of 358 GHz with an antenna configuration with a maximum baseline of 1400 m. The data were correlated with both linear polarisations (XX and YY), with a visibility integration time of 6~s, in four spectral windows, each with 240 channels and 1.875 GHz bandwidth. The spectral windows were centred on 350.8, 352.7, 362.8 and 364.6~GHz, where the last spectral window covers the redshifted rest frequency of the [CII] line ($\nu_{\mathrm{rest}}$ = 1900.5369 GHz). J0519$-$4546 was observed to calibrate the flux density scale and J0522$-$3627 was used to correct the spectral bandpass. J0428$-$5005 was observed as a secondary check source. Phase switching to J0439$-$4522 was carried out at 7-min intervals to calibrate the complex gains resulting from atmosphere-induced phase and amplitude fluctuations. The total on-source integration time was 21 min.\\
We calibrated the raw visibility data using the ALMA pipeline in the Common Astronomy Software Applications CASA package\cite{mcmullin}. The data were then inspected to confirm the quality of the pipeline calibration and that no further flagging was required. Phase-only self-calibration was performed on the continuum with a solution interval of 100 s to correct for residual phase errors. The complex gain corrections from the continuum were applied to the line spectral window. The line data were prepared by fitting a model to the line-free spectral windows and subtracting it from the visibilities, to produce a data set with only the spectral line emission. To improve the signal-to-noise ratio we averaged the data into groups of four velocity channels resulting in twenty-eight independent channels each with a width of 25.7 km\,s$^{-1}$.\\
SPT0418-47 was imaged on a pixel scale of 0.03 arcsec  per pixel with natural weighting of the visibilities and deconvolved using {\sc CLEAN}\cite{hog}. The zeroth-, first- and second- moment maps of the [CII] emission are shown in Figs. \ref{fig1}a-c; we note that these images are intended only for visualisation, as all the modelling and analysis is performed on the visibility data directly. Throughout this study, we assume a $\Lambda$CDM cosmology, with Hubble constant H$_0$ = 67.8 km s$^{-1}$ Mpc$^{-1}$, matter density $\Omega_\mathrm{m}$ = 0.308, and vacuum energy density $\Omega_\mathrm{\Lambda}$ = 0.691\cite{planck}.
\\
\\
\textbf{Lens and kinematic model.} We model the data cube containing the [CII] emission line using a 3D Bayesian lens-kinematic modelling technique that fits the data directly in visibility space (Extended Data Fig. \ref{ed1}). See Rizzo et al.\cite{rizzo2} and Powell et al.\cite{powell} for a more detailed description of the modelling approach, which is briefly summarized below.\\ 
The method used for source reconstruction is grid-based: the background source surface brightness distribution is reconstructed on a triangular adaptive grid\cite{vegetti09} defined by a Delaunay tessellation. The source grid automatically adapts with the lensing magnification, so that there is a high pixel density in the high-magnification regions, resulting in a spatial resolution of $\sim$ 40 pc in the inner regions and $\sim$ 90 pc in the outer regions for this system. The kinematics of the background galaxy are obtained by fitting the lensed data directly in a hierarchical Bayesian fashion, where a 3D kinematic model, $\mathbf{s_{kin}}$, that describes a rotating disk is used as a regularizing prior for a pixellated source reconstruction.\\
Taking advantage of the fact that gravitational lensing conserves surface brightness, and considering the observational noise $\mathbf{n}$, the data in the visibility space $\mathbf{d}$ and the source surface brightness distribution $\mathbf{s}$ can be related to each other via the following set of linear equations:
\begin{equation}
\mathsf{DL}(\boldsymbol{\eta_{\mathrm{lens}}})\mathbf{s} + \mathbf{n} = \mathbf{d} 
\label{eq:lens}
\end{equation}
where $\mathsf{L}$ is the lensing operator, which is a function of the lensing parameters $\boldsymbol{\eta_{\mathrm{lens}}}$, while $\mathsf{D}$ is the non-uniform discrete Fourier transform operator. Because both $\boldsymbol{\eta_{\mathrm{lens}}}$ and $\mathbf{s}$ are unknown, equations (\ref{eq:lens}) can not be simply inverted. We herefore fit the data within the framework of Bayesian statistics by using two levels of inference.\\
At the first level of inference, it can be shown that given the data, a lens-mass model and a source kinematic model, the most probable a posteriori source $\mathbf{s_{MP}}$ is obtained by solving the following set of linear equations:
\begin{equation} \label{eqn:leastsquares}
   \left[(\mathsf{DL)^T C ^{-1} DL} + \lambda \mathsf{R^T R} \right] \mathbf{s} = \mathsf{(DL)^T C}^{-1} \mathbf{d} + \lambda \mathsf{R^T R} \mathbf{s_{kin}}
\end{equation}
where $\mathsf{C}^{-1}$ is the noise covariance and $\mathsf{R}$ is the source regularization form with regularization strength $\lambda$.\\
At the second level of inference, the lens parameters $\boldsymbol{\eta_{\mathrm{lens}}}$, the regularization strength $\lambda$ and the kinematic parameters $\boldsymbol{\eta_{\mathrm{kin}}}$ defining $\mathbf{s_{kin}}$ are obtained by maximizing the posterior probability distribution $P(\mathbf{d} | \boldsymbol{\eta_{\mathrm{lens}}}, \boldsymbol{\eta_{\mathrm{kin}}}, \lambda)$ defined as:
\begin{equation} 
	\begin{split}
    	2\log P(\mathbf{d} | \boldsymbol{\eta_{\mathrm{lens}}}, \boldsymbol{\eta_{\mathrm{kin}}}, \lambda) = -\chi^2 - \lambda (\mathbf{s_{MP}} - \mathbf{s_{kin}})\mathsf{^T R^T R} \, (\mathbf{s_{MP}} - \mathbf{s_{kin}}) - \log \det \mathsf{A}\\+ \log \det(\lambda \mathsf{R^T R})  + \log \det (2\pi \mathsf{C}^{-1}).
	\end{split}
\label{eq:post}
\end{equation}
In equation (\ref{eq:post}), $\chi^2 = (\mathsf{DL}\mathbf{s} - \mathbf{d})\mathsf{^{T} C^{-1}(DL}\mathbf{s} - \mathbf{d})$ and the matrix $\mathsf{A}$ is defined as $\mathsf{[(DL)^T C^{-1} DL + \lambda R^T R]}$.\\
Because of the large number of visibilities per channel, n$_{\rm{vis},i}=163190$, we replace the non-uniform discrete Fourier transform operator $\mathsf{D}$ with a non-uniform Fast Fourier Transform (FFT) operator\cite{greengard2004, beatty2005} (NUFFT), which first uses a gridding kernel to interpolate the visibilities onto a regular gridded uv-plane, then applies an FFT, and finally uses an apodization correction to remap the model and the data onto the original uv-sampling. Because the non-uniform FFT does not have an explicit matrix representation, the source inversion, equation (\ref{eqn:leastsquares}), is solved using a preconditioned conjugate gradient solver. We note that for the conjugate gradient method we use a tolerance of 10$^{-12}$, which results in typical accuracies of 10$^{-6}$ for the source inversions.\\
\\
As usually done when modelling galaxy-galaxy strong lensing observations, the lens is described by a projected mass density profile with a cored elliptical power-law distribution plus the contribution of an external shear component of strength $\Gamma_{\mathrm{sh}}$ and position angle $\theta_{\mathrm{sh}}$ (see Extended Data Table \ref{tab1}). This assumption has been shown to provide a good fit to large samples of known lenses, as discussed, for example, by Koopmans et al.\cite{koop} and Barnabé et al.\cite{barnabe}. The dimensionless projected mass density profile is defined as
\begin{equation}
	\kappa\left(x,y\right)=\frac{\kappa_0\left(2-\frac{\gamma}{2}\right)q^{\gamma-\frac{3}{2}}}{2\left[q^2\left(x^2+r_{\mathrm{c}}^2\right)+y^2\right]^{\frac{\gamma-1}{2}}}\,,
	\label{eq:pl}
\end{equation}
where $\kappa_0$ is the normalization, $q$ is the projected flattening, $\gamma$ is the density slope (a value of $\gamma = 2$ corresponds to an isothermal profile), $x_0$ and $y_0$ define the center of the mass distribution, $r_{\mathrm{c}}$ is the core radius and $\theta$ is the position angle of the major axis. The lens mass model parameters are in agreement within 2 s.d. with those derived from modelling the dust continuum\cite{spilker}. \\
\\
The kinematic model $\mathbf{s_{kin}}$ is defined by the parameters describing the rotation velocity, the velocity dispersion profile and the geometry of the galaxy. We choose the multi-parameter function
\begin{equation}
	V_{\mathrm{rot}}\left(R\right) = V_{\mathrm{t}}\frac{\left(1+\frac{R_{\mathrm{t}}}{R}\right)^{\beta}}{\left[1+\left(\frac{R_{\mathrm{t}}}{R}\right)^{\xi}\right]^{1/\xi}}
	\label{eq:multi}
\end{equation}
as the functional form for the rotation velocity (see Extended Data Table \ref{tab1}). In equation (\ref{eq:multi}) $V_{\mathrm{t}}$ is the velocity scale, $R_{\mathrm{t}}$ is the turnover radius between the inner rising and outer part of the curve, $\beta$ specifies the power-law behaviour of the curve at large radii and $\xi$ defines the sharpness of the turnover. We choose the above multi-parameter function because it is flexible enough to reproduce the large variety of observed rotation curves, and therefore allows much more freedom than other typically used functions, for example the arctangent.\\
The velocity dispersion profile, $\sigma(R)$ is described by an exponential function, which is more flexible than the more commonplace choice of a constant value\cite{wisni, turner}. We also tested a linear function but found it to be significantly less favoured by the data, by a Bayes factor of 1.8 relative to the exponential model.\\
For the geometry of the kinematic model, defined by the inclination ($i$) and position angle ($PA$), we assume that there is no radial variation (Extended Data Table \ref{tab1}). The surface density of the gas is not a free parameter; instead, we impose a pixel-by-pixel normalization, which is given by the surface brightness distribution obtained from the lens modelling of the zeroth-moment map. The advantage of using this normalization is that it allows us to account for possible asymmetries in the gas distribution. The derivation of the lens mass model and the source kinematics is done using a four-step optimisation scheme, and the uncertainties on the parameters are obtained from the posterior distributions calculated with {\sc MULTINEST}\cite{feroz} (Extended Data Fig. \ref{ed2}), by adopting the user-defined tolerance, sampling efficiency and live points of 0.5, 0.8 and 200. We then verified that the evidences estimated by {\sc MULTINEST} flatten as a function of the prior volume.
As described in our method paper\cite{rizzo2}, for each parameter, we adopt priors that are flat in the intervals $[\boldsymbol{\overline{\eta}_{\mathrm{lens/kin}}} - 0.2 \boldsymbol{\overline{\eta}_{\mathrm{lens/kin}}}, \boldsymbol{\overline{\eta}_{\mathrm{lens/kin}}} + 0.2 \boldsymbol{\overline{\eta}_{\mathrm{lens/kin}}}]$, where $\boldsymbol{\overline{\eta}_{\mathrm{lens/kin}}}$  are the best-fitting parameters, inferred from the nonlinear optimization. To be conservative, we report as errors in the parameters the sum in quadrature of the following two contributions: the 1 s.d. uncertainty on the posterior distributions derived by {\sc MULTINEST}and the difference between the maximum a posteriori parameter values obtained by {\sc MULTINEST} and by the non-linear optimizer. These values of the uncertainties are in line with what expected from the tests presented in Rizzo et al.\cite{rizzo2}, where we applied our methodology to data characterized by spatial and spectral resolutions of at least a factor of three worse than those analysed in this manuscript.\\
\\
\textbf{Direct Forward modelling} For a consistency check, we also derived the lens ($\boldsymbol{\eta_{\mathrm{lens}}}$) and kinematic parameters ($\boldsymbol{\eta_{\mathrm{kin}}}$) by using a direct forward modelling methodology. The main difference to the methodology described in Rizzo et al.\cite{rizzo2} and summarized above is that a parametric, rather than a pixellated source is used to fit the data. The parametric source is defined by the same 3D kinematic model, $\mathbf{s_{kin}}$, used as a regularizing prior for the source in our fiducial method. In this implementation, to marginalize over $\boldsymbol{\eta_{\mathrm{lens}}}$ and $\boldsymbol{\eta_{\mathrm{kin}}}$, we maximize the posterior probability distribution (see equation (\ref{eq:post}) for comparison), defined as $2\log P(\mathbf{d}|\boldsymbol{\eta_{\mathrm{lens}}}, \boldsymbol{\eta_{\mathrm{kin}}}) = - (\mathsf{DL}\mathbf{s_{kin}} - \mathbf{d})\mathsf{^{T} C^{-1}(DL}\mathbf{s_{kin}} - \mathbf{d})$. \\
The parameters obtained with the forward modelling method are within the 1 s.d. uncertainties with respect to those obtained with our fiducial model (see Extended Data Fig. \ref{ed2}).\\
In the direct forward modelling method, the source is forced to have a regular configuration, chosen a priori using the assumptions on the kinematic parametrizations. As explained in detail in Rizzo et al.\cite{rizzo2}, by assuming our fiducial methodology, we instead use the kinematic model as a prior for the source reconstruction. The main advantage of the fiducial approach is that it does not directly impose a parametric model which may not be a good fit to the data.\\
\\
\textbf{Beam smearing correction.} In recent years, observational studies have used the $V/\sigma$ ratio to study the kinematic evolution of galaxy populations and distinguish between rotation- and dispersion-dominated systems. However, the measurement of the $V/\sigma$ ratio is a challenge for high-redshift, marginally resolved studies because of the so-called beam-smearing effect\cite{edt1}. This effect can cause a strong overestimation of the velocity dispersion and underestimation of the rotation velocity. For galaxies in the redshift range 1 to 3, there is no consensus whether or not there is an increase of the $V/\sigma$ ratio with time\cite{edt, wisni, lelli, turner, harrison, swin17}. For galaxies at $z \gtrsim 4$ there have been very few attempts to go beyond the fitting of the integrated line profile\cite{jones, smit}. However, marginally resolved observations and a low signal-to-noise ratio have prevented a measurement of a beam-smearing-corrected $V/\sigma$ ratio. The combination of the high signal-to-noise ratio and the spatial resolution that is achievable thanks to gravitational lensing and to an intrinsically beam-smearing corrected kinematic analysis has allowed us to probe the intrinsic $V/\sigma$ ratio for the first time at this high redshift. Different definitions of the $V/\sigma$ ratio can be found in literature. Here we use of the two most popular ones (see Extended Data Table \ref{tab2}). We obtain $V/\sigma = 9.7\pm0.4$ when we calculate $V/\sigma$ as the ratio between the maximum rotation velocity $V_{\mathrm{max}}$ and the mean velocity dispersion $\sigma_{\mathrm{m}}$, whereas we get $V/\sigma = 13.7\pm0.7$, when we calculate it as the ratio between the flat part of the rotation velocity $V_{\mathrm{flat}}$ and the velocity dispersion 
at outer radii ($R > 1$ kpc) $\sigma_{\mathrm{ext}}$. The values of $V/\sigma$ for SPT0418--47 seem to indicate that there is no decrease of $V/\sigma$ with time. However, larger samples will be needed to confirm this trend. \\
\\
\textbf{$\mathbf{V/\sigma}$ for the comparison samples.} To determine the evolution of the dynamical properties of galaxies across cosmic time, different studies have used different gas tracers (see column 2 of Extended Data Table \ref{tab3}) and different extraction methods to obtain their physical parameters. The comparison of $V/\sigma$ for SPT0418--47 with values derived from different gas tracers is based on the assumption that the ionized gas kinematics reflects the cold gas motions of galaxies, as observed for a few galaxies\cite{ubler18, girard}. However, by comparing the results obtained from different tracers and with different techniques, a recent study\cite{ubler19} found that velocity dispersions measured from ionized tracers tend to be $\sim$ 10 - 15 km s$^{-1}$ higher than those measured from molecular tracers. If we add this 15 km s$^{-1}$ to our inferred [CII] velocity dispersions, we find $V_{\mathrm{max}}/\sigma_{\mathrm{m}} = 6.6$ and $V_{\mathrm{flat}}/\sigma_{\mathrm{ext}} = 7.8$, which are still much higher than those observed and those predicted by galaxy evolution models. We also stress that whereas the above analysis mostly compared the molecular and the ionized gas tracers, in this study we focus on the [CII] emission, which traces both cold and warm gas. In Extended Data Table \ref{tab3}, we show the main assumptions made to extract the velocities and velocity dispersions that enter the calculation of the $V/\sigma$ ratio for the samples shown in Fig. \ref{fig3}. \\
\\
\textbf{$\mathbf{V/\sigma}$ in galaxy evolution models.} There is a general consensus\cite{hung, teklu, bird, martizzi} that galaxy disks at high redshift are much more turbulent than their local counterparts in all components (stars, warm and cold gas).\\
In Fig. \ref{fig3} we show the $V/\sigma$ predicted (or assumed) by current galaxy evolution models:
\begin{itemize}
\item The light-blue area shows the predicted evolution of $V/\sigma$ in the TNG50 simulation\cite{pillepich}, where the gas kinematics is derived from H$\alpha$ emitting gas. However, because, by construction, the ionized and molecular gas of TNG50 galaxies have the same dynamics, we can compare these estimations with our measurement coming from the [CII] emission line. TNG50 galaxies at $z \sim 4$ have median $V_{\mathrm{max}}/\sigma_{\mathrm{m}} \sim 3$, with a standard deviation of 1.5, which means that SPT0418--47 is 4.5 standard deviations away from the median value.
\item By using an analytic approach and a cosmological mesh-refinement simulation, two recent studies\cite{dekel, zolotov15} found that galaxies at $z \gtrsim$ 3 are dominated by violent disk instabilities, which lead to $V/\sigma$ values $\lesssim2$ in all components (grey area in Fig. \ref{fig3}).
\item By using an analytic model that combines stellar feedback and gravitational instabilities, Krumholz et al.\cite{krum} derived a prediction for the correlation between cold gas velocity dispersion and SFR, such that SFR = $(0.42 f_{\mathrm{eff, gas}} V^{2} \sigma)/(\pi G Q_{\mathrm{min}})$ (their equation 60), where we have used the constants appropriate for high-redshift galaxies\cite{krum, ubler19}. By using $f_{\mathrm{eff, gas}} = f_{\mathrm{gas}} \times 1.5 =  0.53 \times 1.5$, SFR = 352 M$_\odot$/yr, $V = 259$ km s$^{-1}$, $Q_{\mathrm{min}} = 0.88$, we derived a value of $\sigma = 183_{-42}^{+53}$ km s$^{-1}$, which is a factor of $\sim 6$ higher than our measured value of 32 km s$^{-1}$, and implies $V/\sigma = 1.4_{-0.3}^{+0.4}$ (black band in Fig. \ref{fig3}).
\item Hayward and Hopkins\cite{hay} used an analytic model to study the effects of stellar feedback in regulating star formation and driving outflow. In this case, the gas velocity dispersion is not a predicted quantity, but it is derived by assuming the following relation between $\sigma$ and the circular velocity: $\sigma \sim f_{\mathrm{gas}} V_{\mathrm{c}}/\sqrt{2}$. For the values of the gas fraction and $V_{\mathrm{c}}$ measured for SPT0418--47 by our analysis, this analytical model implies a value of $\sigma$ = 120 km s$^{-1}$, which is a factor of $\sim4$ higher than the measured value, resulting in a $V/\sigma = 2.6$. In Fig. \ref{fig3} (green band), we show the predictions of $V/\sigma$ for this model, obtained by using the relations of ref.\cite{hay} for velocity versus stellar mass and the redshift evolution of velocity dispersion with respect to the gas fraction.
\end{itemize}  
\textbf{Dynamical model and Toomre parameter.} Under the assumption that the total galactic gravitational potential $\Phi$ is axisymmetric, the rotation velocity  $V_{\mathrm{rot}}(R)$ of the gas, in cylindric coordinates ($R$, $\phi$, $z$), is related to $\Phi$ by the equation $R (\partial \Phi/ \partial R)_{z=0} = V_{\mathrm{c}}^2 =  V_{\mathrm{rot}}^2 + V_{\mathrm{A}}^2$, where $V_{\mathrm{c}}$ is the circular velocity, and $V_{\mathrm{A}}$ is the asymmetric-drift correction that accounts for the pressure support due to the random motions. Under the assumptions that the gas of the rotating disk has a thickness that is independent of the radius\cite{iorio}, it is thin and it has a spatial distribution described by an exponential profile, $\Sigma_{\mathrm{gas}} = \Sigma_{0} \exp{ (-R/R_{\mathrm{gas}})}$, the expression for $V_{\mathrm{A}}$ is given by
\begin{equation}
	V^2_{A} = -R \sigma^2 \partial \ln (\sigma^2 \Sigma_{\mathrm{gas}})/ \partial R = -R \sigma^2 \partial \ln (\sigma^2 \exp(-R/R_{\mathrm{gas}}))/ \partial R.
	\label{eq:va}
\end{equation}
To measure the scale radius $R_{\mathrm{gas}}$ we divide the zeroth-moment map of the reconstructed source (Fig. 1d) into rings (with centres, $PA$ and $i$ defined by the values of the kinematic model; see Extended Data Table \ref{tab1}) and we calculate the surface densities at a certain radius as azimuthal averages inside that ring. The surface density profile obtained in this way is then fitted using the exponential profile, resulting in a value of $R_{\mathrm{gas}} = 0.9$ kpc. The resulting asymmetric-drift correction, equation (\ref{eq:va}), gives a small contribution ($\lesssim 1\%$) with respect to $V_{\mathrm{rot}}(R)$.\\
\\
To derive the contribution of the gas, stellar and dark-matter components to the total gravitational potential, we model the circular velocity as $V_{\mathrm{c}} = \sqrt{V^2_{\mathrm{star}} +V^2_{\mathrm{gas}}+ V^2_{\mathrm{DM}}}$. Here $V_{\mathrm{star}}$ is the stellar contribution to the circular velocity, under the assumption that this component is described by a Sérsic profile\cite{limaneto, terzic}, defined by the total stellar mass $M_{\mathrm{star}}$, the effective radius radius $R_{\mathrm{eff}}$ and the Sérsic index $n$. $V_{\mathrm{gas}}$ is the gas contribution, under the assumption that the gas in this galaxy has a distribution described by an exponential profile, as traced by the [CII] emission line. The scale length that enters in $V_{\mathrm{gas}}$ is fixed at the value $R_{\mathrm{gas}}$ found in the previous paragraph, and the only free parameter of the fit for $V_{\mathrm{gas}}$ is the conversion factor ($\alpha_{\mathrm{[CII]}}$) between the total [CII] luminosity and the total gas mass. A number of recent studies have found that [CII] is a good tracer of the total gas mass\cite{zanella, gull}. For SPT0418--47, the [CII] luminosity is 1.8 $\times 10^{9}$~L$_{\odot}$, obtained by computing the zeroth-moment map of the [CII] emission as the signal integrated along the spectral axis at each pixel of the reconstructed source. The dark-matter contribution $V_{\mathrm{DM}}$ is modelled as a Navarro-Frenk-White\cite{navarro} (NFW) spherical halo, with a concentration parameter of $c = 3.06$. The latter is obtained by averaging the values of $c$ at $z = 4.22$ for dark matter haloes with masses between $10^{10} M_{\odot}$ and $10^{13} M_{\odot}$, assuming the mass-concentration relation estimated in N-body cosmological simulations\cite{dutton}. We notice that at this redshift, $c$ is almost independent of the dark-matter halo mass, varying by just 6 per cent for a variation of 3 orders of magnitude in the halo mass. To test the effect of our assumption of a constant concentration, we repeated the analysis with a $c$ that is free to vary according to either a uniform prior or a Gaussian prior centred on the predicted mass-concentration relation, and found that all inferred dynamical parameters do not change significantly with $c$.\\
\\
We summarize all our assumptions for the dynamical model in the second column of Extended Data Table \ref{tab4}, and the corresponding free parameters are shown in the third column of the same table. We compute the Bayesian posterior distribution of these parameters using {\sc DYNESTY}, a python implementation of the Dynamic Nested Sampling algorithm\cite{dyn} (see Extended Data Fig. \ref{ed3} and Table \ref{tab5}). We use log-uniform priors for the mass parameters and uniform priors for the scale radii (fourth column of Extended Data  Table \ref{tab4}). The inferred $M_{\mathrm{star}}=1.2^{+0.2}_{-0.1} \times 10^{10}$ M$_\odot$ is in excellent agreement with the value of ($9.5 \pm 3.0) \times 10^{9}$ M$_\odot$ found in a recent independent study\cite{deb} by fitting the spectral energy distribution of this galaxy. For $\alpha_{\mathrm{[CII]}}$ we employ a uniform prior in the range corresponding to $\pm3$ standard deviations around the mean value of 30~M$_{\odot}$/L$_{\odot}$, derived from a large sample of low- and high-redshift galaxies\cite{zanella}. We infer a value of $\alpha_{\mathrm{[CII]}}=7.3^{+1.0}_{-1.2}$~M$_{\odot}$/L$_{\odot}$, in agreement with other studies of z $\sim$ 4 dusty starburst galaxies\cite{gull}.\\
\\
In Extended Data Table \ref{tab5} we show some relevant physical quantities (see also Fig. \ref{fig4}, panel d) derived from our dynamical analysis. In particular, for each of these we quote the 16th, 50th and 84th percentile, which were obtained from the full posterior sample points returned by {\sc DYNESTY}.\\
\\ 
We also calculate the Toomre parameter\cite{toomre} (Fig. \ref{fig2}d), using the general definition $Q(R) = (\sigma \kappa )/ (\pi G \Sigma_{\mathrm{gas}})$. Here, $\kappa$ is the epicycle frequency defined as $\kappa = \sqrt{ R d\Omega^2/dR + 4\Omega^2}$, where $\Omega = V_{\mathrm{rot}}/R$ is the angular frequency.\\
\\
\textbf{Comparison with local ETGs.} In Fig. \ref{fig4} we show some physical quantities for local ETGs taken from multiple studies of the ATLAS$^{\mathrm{3D}}$ survey\cite{cappellari, mcdermid, kraj, cappellari2013a, cappellari2013b, emsellem}. For comparison with SPT0418--47, we show for each quantity an upper limit, calculated under the assumption that all gas is converted into stars, and the observed quantities (orange diamonds), which can also be interpreted as a lower limit if all the gas that we observe will be ejected by feedback processes. However, we caution that this upper limit does not take into account the accretion of gas if no baryons are expelled by outflows. In this case, the red diamond could move further towards the upper right corner in the size-mass plane (Fig. \ref{fig4}a).\\
\\
\textbf{Dust continuum and SFR.}
Using the parameters of the lens mass model shown in Extended Data Table \ref{tab1}, we perform a pixellated source reconstruction of the dust continuum at 160 $\mu$m (rest frame), which allows us to calculate a total magnification factor $\mu$ of 32.3$\pm 2.5$. SPT0418--47 has an observed (unlensed) infrared luminosity of $L_{\mathrm{IR, obs}} = (7.7\pm1.3) \times 10^{13}$~L$_{\odot}$\cite{aravena}. This value was obtained from a spectral energy distribution fitting of seven photometric data points in the wavelength rest frame range between 48 and 574 $\mu$m. Under the assumption that the morphology of the source is the same for all photometric points in the infrared band, we can use the magnification factor estimated from the emission at 160 $\mu$m to estimate an intrinsic luminosity of $L_{\mathrm{IR}} = L_{\mathrm{IR, obs}}/\mu = (2.4\pm0.4) \times 10^{12}$~L$_{\odot}$. By assuming that the infrared emission comes from the thermal emission of dust, heated by the radiation field coming from young stars, we compute SFR = ($352\pm65$) ~M$_{\odot}$~yr$^{-1}$. This value is derived from $L_{\mathrm{IR}}$, by applying a conversion factor of $1.48 \times 10^{-10}$~M$_{\odot}$~(yr~L$_{\odot}$)$^{-1}$, which is valid for a Kroupa Initial Mass Function\cite{kennicutt}. The gas-depletion timescale is $M_{\mathrm{gas}}$/SFR $= 38\pm9$~Myr.\\
\\
\textbf{Velocity dispersion in starburst galaxies.} By comparing the gas kinetic energy and the energy injected by stellar feedback, we determine whether the turbulence driven by supernova explosions can explain the velocity dispersion measured for SPT0418--47. The gas kinetic energy is given by\cite{tamburro, utomo}: $E_{\mathrm{kin}} = 3/2 \,M_{\mathrm{gas}}\,\sigma_{\mathrm{m}}^2 \sim 5 \times 10^{56}$ (see Extended Data Table \ref{tab5}). We estimate the energy injected by supernova explosions as\cite{tamburro, utomo} $E_{\mathrm{SNe}} = \eta_{\mathrm{SN}}\,\mathrm{SFR}\,\epsilon_{\mathrm{SN}}\,E_{\mathrm{SN}}\,t_{\mathrm{D}}$. Here, $\eta_{\mathrm{SN}}$ is the supernova rate, $\epsilon_{\mathrm{SN}}$ is the SFR efficiency (that is, the fraction of the supernova energy, $E_{\mathrm{SN}} = 10^{51}$ erg, converted into turbulence) and $t_{\mathrm{D}}$ is the dissipation rate of the turbulence. We assume $\eta_{\mathrm{SN}} = 0.01 M_{\odot}^{-1}$ \cite{tamburro}. For the highly uncertain parameter $\epsilon_{\mathrm{SN}}$ we assume an upper limit value of 0.1, based on high-redshift observations\cite{rafelski}. $t_{\mathrm{D}}$ is calculated as $t_{\mathrm{D}} = h/\sigma_{\mathrm{m}}$ where $h$ is the scale height of the gas, assumed to be $\sim 500$ pc. Under this assumption, the estimate $E_{\mathrm{SNe}}$ is $\sim 5 \times 10^{56}$ erg, in agreement with $E_{\mathrm{kin}}$, suggesting that the velocity dispersion measured for SPT0418--47 can be explained as being produced by turbulence driven by stellar feedback. Furthermore, we note that starburst galaxies both at low\cite{leroy, lelli14} and high redshifts\cite{lelli} have values of velocity dispersions, measured from molecular or neutral gas, that are comparable with those found for SPT0418--47.
\end{methods}

\newpage
\renewcommand{\tablename}{Extended Data Table}
\begin{table}[h!]
  \centering
  \caption{\small{\textbf{Lens and source kinematic parameters.} The lens parameters describe a projected mass density profile with a cored elliptical power-law distribution, equation \eqref{eq:pl}, plus the contribution of an external shear component ($\Gamma_{\mathrm{sh}}$, $\theta_{\mathrm{sh}}$). The kinematic parameters describe a rotating disk with a rotation curve defined by a multi-parameter function, equation \eqref{eq:multi}, a velocity dispersion profile defined by an exponential function ($\sigma_{0}$, $R_{0}$), and a geometry defined by the inclination ($i$) and the position angles ($PA$).
}}
  \medskip
  \begin{tabular}{c c |c c}
    \hline
    \noalign{\medskip}
    \multicolumn{2}{c}{Lens parameters} & \multicolumn{2}{c}{Kinematic parameters} \\
	\noalign{\medskip}
	\hline
	\noalign{\smallskip}
    $\kappa_0$ (arcsec) & $1.22\pm0.04$ 			&	$i$ ($^{\circ}$) & $54\pm$2\\
    $\theta$ ($^{\circ}$) & $22.6\pm1.6$   			&	$PA$ ($^{\circ}$) & $-27\pm$3\\
    $q$ & $0.91\pm0.02$  						&	$V_{\mathrm{t}}$ (km\,s$^{-1}$) & $245.1\pm$0.8\\
    $\gamma$ & $2.05\pm0.09$				&	$R_{\mathrm{t}}$ (kpc) & $0.14\pm$0.03\\
    $\Gamma_{\mathrm{sh}}$ & 	$0.0082\pm0.0003$ 	&	$\beta$ & $0.80\pm$0.02\\
    $\theta_{\mathrm{sh}}$ ($^{\circ}$) & $29.9\pm3.0$ 	&	$\xi$ & $2.0\pm$0.4\\
     &							&	$\sigma_{0}$ (km\,s$^{-1}$) & $58.1\pm$1.8\\
     &							&	$R_{0}$ (kpc) & $1.7\pm$0.1\\     
    \hline
    \end{tabular}
    \label{tab1}
\end{table}

\begin{table}[h!]
	\centering
	\caption{\small{\textbf{Kinematic properties for SPT0418-47 derived under different assumptions.} Parameters representing the rotation velocities and velocity dispersion profile, as well as the rotation support for this galaxy, calculated under different definitions. $^{a}$Maximum rotation velocity. $^{b}$Mean velocity dispersion. $^{c}$Rotation-to-random-motions ratio, calculated from $V_{\mathrm{max}}$ and $\sigma_{\mathrm{m}}$. $^{d}$Flat rotation velocity, calculated using the flat part of the rotation curve (R $> 2$ kpc, see Fig. \ref{fig2}c). $^{e}$Velocity dispersion at outer radii (R $> 1$ kpc). $^{f}$Rotation-to-random-motion ratio, calculated from $V_{\mathrm{flat}}$ and $\sigma_{\mathrm{ext}}$.}}
	\medskip
	\small{
	\begin{tabular}{c c c c} 
		\hline
		\noalign{\medskip}
		\multicolumn{4}{c} {Global kinematic properties}\\
		\noalign{\medskip}
		\hline
		\noalign{\smallskip}
		$^{a} V_{\mathrm{max}}$ (km\,s$^{-1}$) & 308$\pm$4  & $^{d} V_{\mathrm{flat}}$ (km\,s$^{-1}$) & 259$\pm$1 \\
		$^{b} \sigma_{\mathrm{m}}$ (km\,s$^{-1}$) & 32$\pm$1& $^{e} \sigma_{\mathrm{ext}}$ (km\,s$^{-1}$) & 18$\pm$1\\
		$^{c} V_{\mathrm{max}}/\sigma_{\mathrm{m}}$ & 9.7$\pm$0.4 & $^{f} V_{\mathrm{flat}}/\sigma_{\mathrm{ext}}$ & 13.5$\pm$0.7\\
		\hline
	\end{tabular}
	}
	\label{tab2}
\end{table}

\begin{table}[ht!]
  \centering
  \caption{\small{\textbf{Kinematic measurements for the comparison samples.} $^{a}$The value of $\sigma$ are not provided and we assume a value of 10 km s$^{-1}$, typical of HI in local spiral galaxies$^{11}$. $^{b}$ $V_{\mathrm{PV, max}}$ and $V_{\mathrm{PV, min}}$ are the maximum and minimum velocity along the position-velocity diagram.}}
  \medskip
  \small{
  \begin{tabular}{c c c c} 
    \hline
    \noalign{\medskip}
    Study & Tracer & $V$ & $\sigma$ \\
	\noalign{\medskip}
	\hline
	\noalign{\smallskip}
    Lelli et al.\cite{lelli16} & HI & Flat part of the rotation curve ($V_{\mathrm{flat}}$) & $^a$10 km/s\\
    Swinbank et al.\cite{swin17}  & H$\alpha$, [OII]  & Extracted at 3$R_{\mathrm{disk}}$ & Average ($\sigma_{\mathrm{m}}$)\\
    Harrison et al.\cite{harrison}  & H$\alpha$ & Extracted at 3.4$R_{\mathrm{disk}}$ & Median value at $R > 3.4\,R_{\mathrm{disk}}$\\
    Di Teodoro et al.\cite{edt}  & H$\alpha$ & Flat part of the rotation curve ($V_{\mathrm{flat}}$) & Average ($\sigma_{\mathrm{m}}$)\\
    Wisnioski et al.\cite{wisni}  & H$\alpha$ &  $(V_{\mathrm{PV, max}} - V_{\mathrm{PV, min}})/2$$^b$ & Average from outer regions ($\sigma_{\mathrm{ext}}$)\\
    Lelli et al.\cite{lelli}  & [CI] & Flat part of the rotation curve ($V_{\mathrm{flat}}$) & Average ($\sigma_{\mathrm{m}}$)\\
    Turner et al.\cite{turner} & [OIII] & Extracted at 3.4$R_{\mathrm{disk}}$ & Median\\
    \hline
    \end{tabular}
    }
    \label{tab3}
\end{table}

\begin{table}[ht!]
  \centering
  \caption{\small{\textbf{Assumptions for the dynamical fit}}}
  \medskip
  \begin{tabular}{c c c c}
    \hline
    \noalign{\medskip}
    Component & Density profile & Free parameters & Priors\\
    \noalign{\medskip}
    \hline
    \noalign{\smallskip}
	Stars & Sérsic &  $M_{\mathrm{star}}$ & [$10^7, 10^{11}$] $M_{\odot}$\\
	 &  &  $R_{\mathrm{e}}$ & [$0.04, 2.0$] kpc\\
	& &   $n$ & [0.5, 10] \\
	Gas & Exponential &  $\alpha_{\mathrm{[CII]}}$ & [$3.8, 238.0$] $M_{\odot}/L_{\odot}$\\
	DM & NFW &  $M_{\mathrm{DM}}$ & [$10^{10}, 10^{13}$] $M_{\odot}$\\
    \hline
    \end{tabular}
    \label{tab4}
\end{table}

\begin{table}[h!]
	\centering
	\caption{\small{\textbf{Physical quantities for SPT0418-47 derived from the kinematic and dynamical modelling.} Left: parameters inferred from a dynamical fit to the circular velocity. The stellar component is described by a Sérsic profile, the gas disk by an exponential profile and the dark matter is defined by a NFW profile. Right: the quantities in square brackets are calculated considering the gas component, under the assumption that all the gas that we observe today will be converted into stars, preserving the disk configuration. $^{a}$Total gas mass, computed as $M_{\mathrm{gas}} = \alpha_{\mathrm{[CII]}} L_{\mathrm{[CII]}}$. $^{b}$Total baryonic mass, computed as $M_{\mathrm{bar}} =  M_{\mathrm{star}} + M_{\mathrm{gas}}$. $^{c}$Baryonic half mass radius. $^{d}$Dark matter fraction within the half mass radius. $^{e}$Stellar-to-halo mass ratio: $M_{\mathrm{star}}/M_{\mathrm{DM}}$ [($M_{\mathrm{star}} + M_{\mathrm{gas}})/M_{\mathrm{DM}}$]. $^{f}$Gas fraction: $M_{\mathrm{gas}}/(M_{\mathrm{star}} + M_{\mathrm{gas}})$. $^{g}$Virial velocity of the dark-matter halo. $^{h}$Virial radius of the dark-matter halo. $^{i}$Maximum velocity for an NFW halo, computed as  $V_{200} \sqrt{0.216\,c/ (\ln(1+c)-c/(1+c))}$.}}
	\medskip
	\small{
	\begin{tabular}{c c | c c} 
		\hline
		\noalign{\medskip}
		\multicolumn{2}{c} {Parameters of the dynamical model} & \multicolumn{2}{c} {Derived parameters}\\
		\noalign{\medskip}
		\hline
		\noalign{\smallskip}
		$M_{\mathrm{star}}$ ($10^{10}$$M_{\odot}$) & $1.2^{+0.2}_{-0.1}$ & $^{a} M_{\mathrm{gas}}$ ($10^{10}$$M_{\odot}$) & $1.3^{+0.2}_{-0.2}$\\
		$R_{\mathrm{e}}$ (kpc) & $0.22^{+0.04}_{-0.02}$ & $^{b} M_{\mathrm{bar}}$ ($10^{10}$$M_{\odot}$) & $2.5^{+0.2}_{-0.1}$\\
		$n$  & $2.2^{+0.3}_{-0.2}$ & $^{c} R_{\mathrm{e, bar}}$ (kpc) &  [$0.75^{+0.06}_{-0.06}$]\\
		$M_{\mathrm{DM}}$ ($10^{12}$$M_{\odot}$)& $1.7^{+0.3}_{-0.3}$ & $^{d} f_{\mathrm{DM}}(<R_{\mathrm{e}})$ & $0.018^{+0.005}_{-0.003}$ [$0.095^{+0.008}_{-0.007}$]\\	
		$\alpha_{\mathrm{[CII]}}$  ($M_{\odot}$/$L_{\odot}$) & $7.3^{+1.0}_{-1.2}$ & $^{e} f_{\star}$ ($10^{-3}$) & $7.1^{+1.0}_{-0.8}$ [$14.9^{+3.7}_{-2.6}$] \\
		& & $^{f} f_{\mathrm{gas}}$ & $0.53^{+0.06}_{-0.08}$\\
		& & $^{g} V_{200}$ (km\,s$^{-1}$)& $320^{+17}_{-18}$ \\
		& & $^{h} R_{200}$ (kpc) &  $70^{+4}_{-4}$\\
		& & $^{i} V_{\mathrm{max}}$ (km\,s$^{-1}$) & $323^{+18}_{-19}$\\
		\hline
	\end{tabular}
	}
	\label{tab5}
\end{table}

\setcounter{figure}{0}

\begin{Extended Data Figure}[h!]
	\centering
	\includegraphics{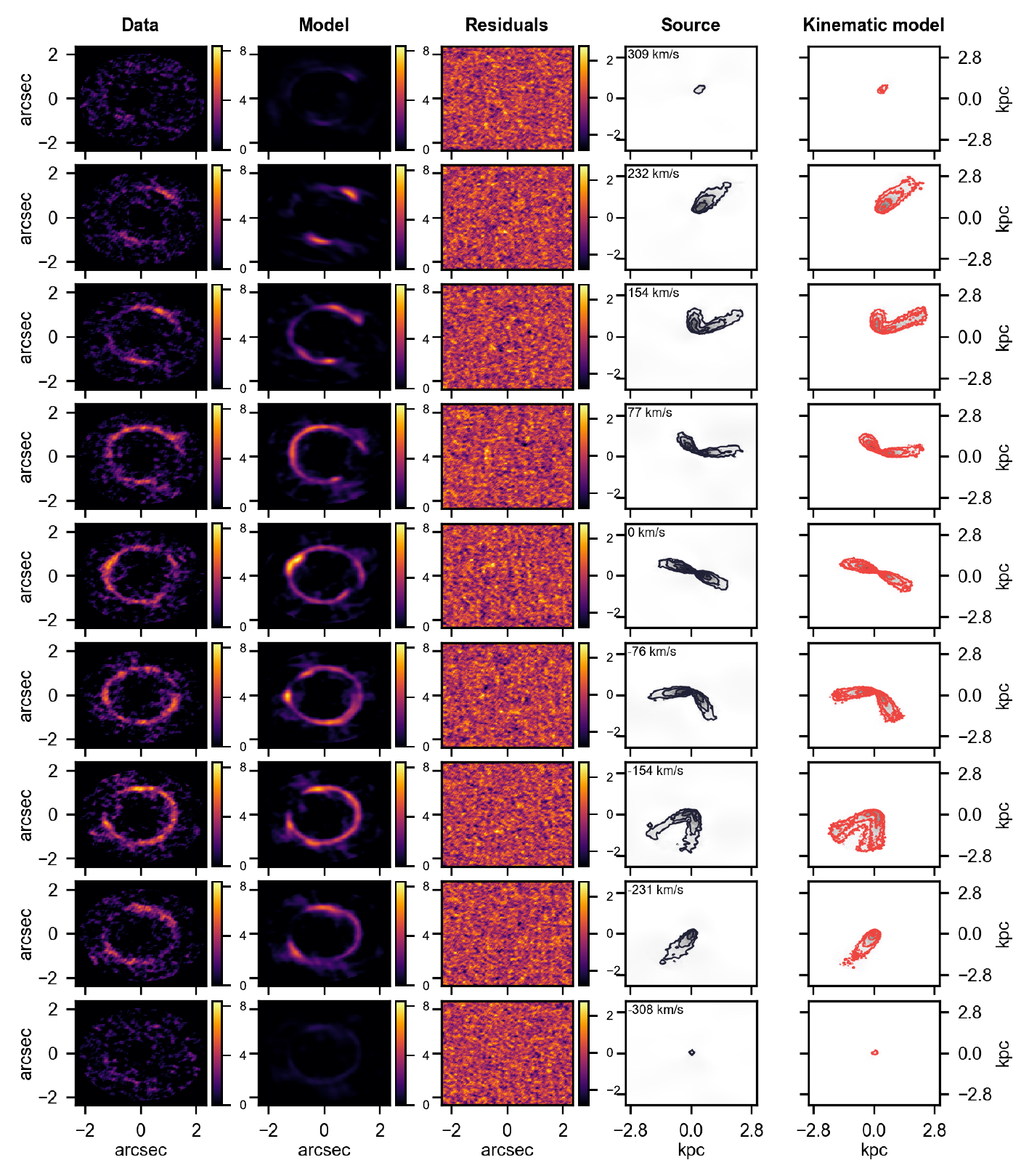}
	\caption{\small{\textbf{Reconstrunction of the [CII] emission and kinematic model.}  The rows show some representative channel maps at the velocity shown on the upper left corner of column 4. Columns 1 and 2 show the dirty image of the data and the model, respectively, colour-coded by the flux in units of mJy/beam. Column 3 shows the residuals (data - model) normalized to the noise. Column 5 and 6 show the contours of the reconstructed source and of the kinematic model used to constrain the source reconstruction. The contour levels in the last columns are set at n = 0.1, 0.2, 0.4, 0.6, 0.8 times the value of the maximum flux of the kinematic model.}}
	\label{ed1}
\end{Extended Data Figure}

\begin{Extended Data Figure}
	\centering
	\includegraphics[width=1.02\textwidth]{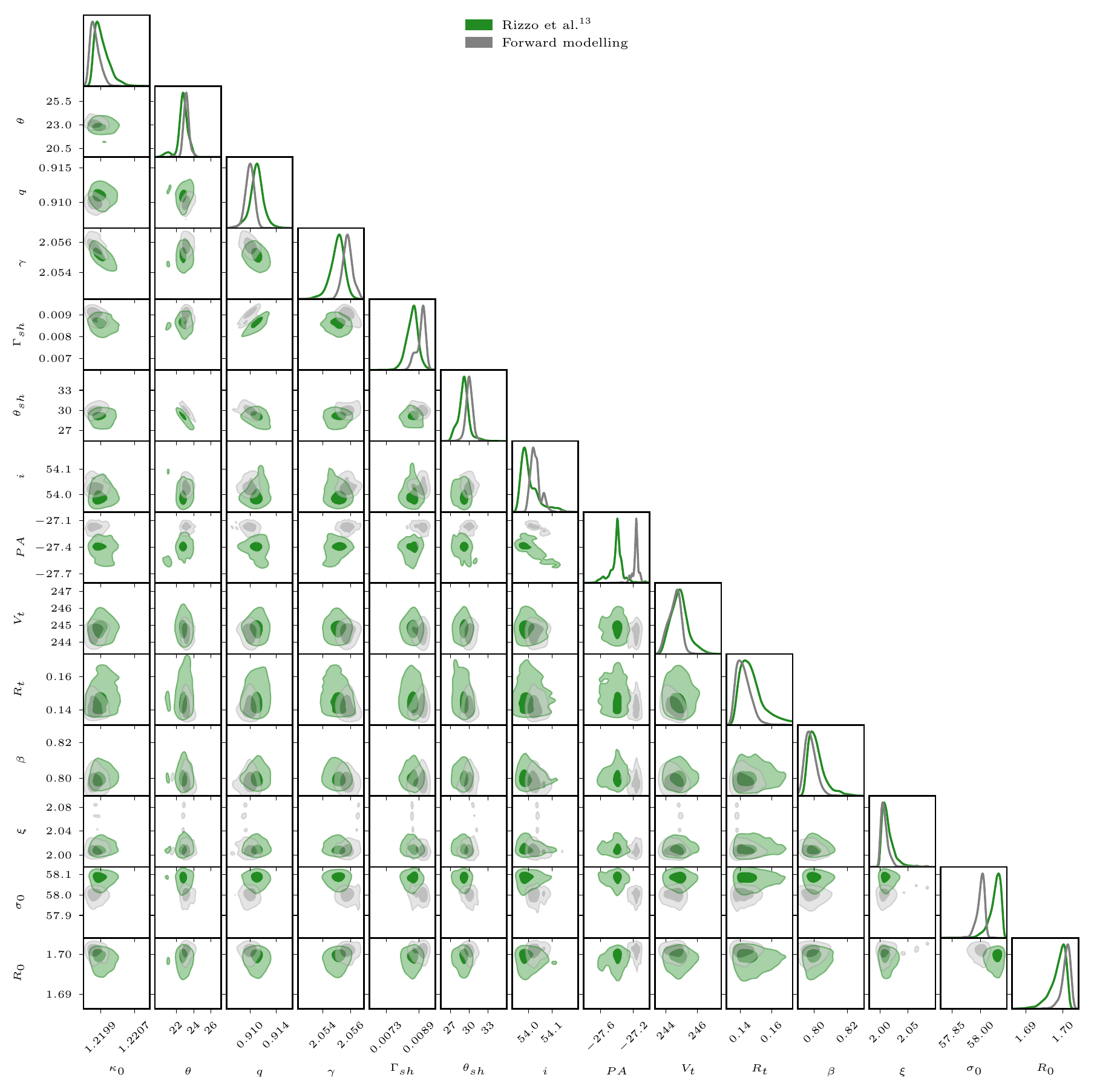}
	\caption{\small{\textbf{Corner plot for the posterior distributions of the lens and kinematic parameters.} The dark and light areas the 2D distributions show the 39\% and 86\% confidence levels, corresponding to 1 s.d. and 2 s.d., respectively, obtained with the fiducial methodology described in Rizzo et al. 2018\cite{rizzo2} (green) and with the direct forward modelling methodology (gray). From left to right, the first six panels show the lens parameters, and the other panel show the source kinematic parameters (see also Extended Data Table \ref{tab1}).}}
	\label{ed2}
\end{Extended Data Figure}

\begin{Extended Data Figure}[h!]
	\centering
	\includegraphics[width=1.02\textwidth]{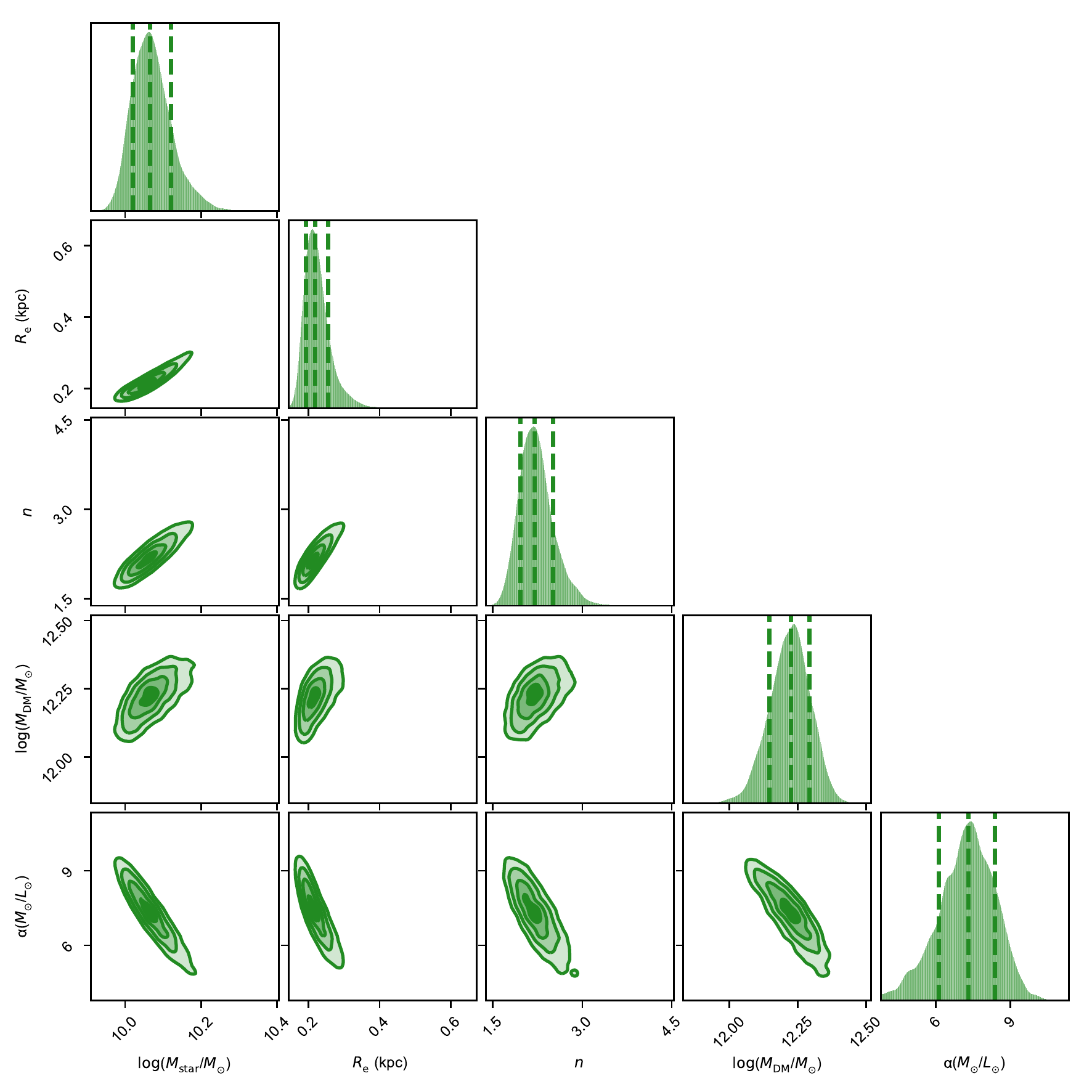}
	\caption{\small{\textbf{Corner plot for the posterior distributions of the dynamical parameters.} The posterior distributions are the results of the decomposition of 	the circular velocity (Fig. \ref{fig2}c) into the physical components contributing to the total gravitational potential: the stars, the gas disk and the dark matter halo. The 		fitted parameters are the stellar mass $M_{\mathrm{star}}$, the effective radius $R_{\mathrm{e}}$, the Sérsic index $n$ (Sérsic profile), the mass of an NFW 		dark-matter halo $M_{\mathrm{DM}}$ and the conversion factor between the [CII] luminosity and the gas mass $\alpha_{\mathrm{[CII]}}$. The dashed lines in 		the 1D histograms show the 16th, 50th and 84th percentiles (see Extended Data Table \ref{tab5}).}}
	\label{ed3}
\end{Extended Data Figure}

\clearpage

\bibliography{reference_spt.bib}

\clearpage
\begin{addendum}
\item [Acknowledgements] This letter makes use of the following ALMA data: 2016.1.01499.S. ALMA is a partnership of ESO (representing its member states), NSF (USA) and NINS (Japan), together with NRC (Canada), NSC and ASIAA (Taiwan), and KASI (Republic of Korea), in cooperation with the Republic of Chile. The Joint ALMA Observatory is operated by ESO, AUI/NRAO and NAOJ. SV has received funding from the European Research Council (ERC) under the European Union's Horizon 2020 research and innovation programme (grant agreement No. 758853). F.R. thanks T. Naab and T. Costa for useful comments and discussions.
\item [Author contributions] F.R., F.F. and S.V. analysed the data. D.P., F.R. and S.V. developed the software used for the lens-kinematic modelling. F.R. developed the software for the dynamical analysis. H.R.S. and F.R. reduced the data. F.R., J.P.M., S.V. and H.R.S. contributed to the writing of the manuscript. F.F. and S.D.M.W. helped with the interpretation of the scientific results. All authors discussed the results and provided comments on the paper.
\item[Author Information] Reprints and permissions information is available at www.nature.com/reprints. The authors declare that they have no
competing financial interests. Correspondence and requests for materials should be addressed to F.R.~(email: frizzo@mpa-garching.mpg.de).
\item [Data Availability] This paper makes use of the following ALMA data: 2016.1.01499.S, available at\\
 \href{http://almascience.eso.org/aq/}{http://almascience.eso.org/aq/}.
\item [Code Availability] The methodology about the source reconstruction and its kinematic modelling is fully explained in Reference 13. The code is not publicly available. However, the reader interested in using this code can contact the first author.
\end{addendum}

\end{spacing}

\end{document}